# The effect of $^{11}$B substitution on the superconductivity in MgCNi$_3$

T. Klimczuk[1,2], M. Avdeev[3], J. D. Jorgensen[3] and R.J. Cava[1]

[1]Department of Chemistry, Princeton University, Princeton NJ 08544, USA

[2]Faculty of Applied Physics and Mathematics, Gdańsk University of Technology,
Narutowicza 11/12, 80-952 Gdańsk, Poland,

[3] Materials Science Division, Argonne National Laboratory, Argonne IL 60439, USA

The crystal structure of boron doped superconducting MgC$_{1-x}$$^{11}$B$_x$Ni$_3$, studied by powder neutron diffraction, is reported. The solubility limit of boron is determined to be approximately x=0.16. The unit cell expands from $a$ = 3.81089(2) Å to 3.81966(2) Å as x increases from x=0 to x=0.155. Boron ($^{11}$B) doping decreases Tc with increasing x: from 7.09K (x=0) to 6.44K (x=0.155).

**Introduction**

The discovery of superconductivity in MgB$_2$ (Ref. 1) motivated a search for new superconducting materials containing light elements like magnesium, boron and carbon. Surprisingly, superconductivity was discovered in MgCNi$_3$ (T$_C$=7K), in which the high proportion of Ni metal suggests that magnetic interactions may result in ferromagnetism rather than superconductivity[2]. Band structure calculations show a narrow peak in the vicinity of the Fermi energy (E$_F$)[3-9]. However, the density of states (DOS) at E$_F$ is not believed be big enough to yield ferromagnetism[6]. The presence of the DOS peak, although with lower intensity, was confirmed by photoemission and x-ray spectroscopy experiments[7,10].

Shim *et al*. showed that the Fermi surface is composed of two bands, and proposed that the strong, narrow DOS peak, located just below E$_F$, corresponds to the $\pi^*$ antibonding state of Ni 3*d* and C 2*p* but with predominant Ni 3*d* character[6]. Therefore, hole doping of MgCNi$_3$ in attempts to move the Fermi level into the DOS peak should be of interest. However, neither increasing T$_C$ nor ferromagnetism was observed by partial substitution of Co, Fe, Ru, or Mn for Ni[11-15] or by inducing carbon deficiency[16,17].

The carbon atom in MgCNi$_3$ plays a critical role in the superconductivity. A single phase superconducting compound occurs only in a narrow range of carbon content (0.85 < x < 1.0)[16]. Shan *et al*. investigated the specific heat of MgC$_x$Ni$_3$ and showed a difference between Sommerfeld parameters of superconducting (x close to 1) and nonsuperconducting (x about 0.85) samples[17]. They proposed that the disappearance of superconductivity is due to a substantial depression of the electron-phonon coupling caused by decreasing x. Recently, a $^{13}$C isotope effect with $\alpha_C$ = 0.54(3) was reported, which indicates that MgCNi$_3$ is predominantly a phonon–mediated superconductor and confirms the important role of carbon in superconductivity[18].

Synthesis of MgCNi$_3$ requires an excess of both Mg and C to compensate for Mg evaporation and to ensure carbon incorporation. Therefore controlled doping of both the Mg-site and the C-site is difficult, and crystal structure analysis is required to determine the true composition. Previous experiments with hole doping on the Ni-site revealed a decrease of T$_C$. However, no attempts have been reported on doping the C-site in MgCNi$_3$. In this case, considering the covalent radii of the elements, there are two obvious candidates for substitution: boron and nitrogen. Here we report a study of superconductivity in MgC$_{1-x}$B$_x$Ni$_3$.

**Experimental**

A series of 0.4g samples with compositions $Mg_{1.2}C_{1.5-x}{}^{11}B_xNi_3$ (x=0, 0.05, 0.1, 0.15, 0.2 and 0.25) was synthesized. The starting materials were bright Mg flakes (99% Aldrich Chemical), fine Ni powder (99.9% Johnson Matthey and Alfa Aesar), glassy carbon spherical powder (Alfa Aesar) and enriched boron metal powder $^{11}B$ (99.5 At.% $^{11}B$ – Eagle-Picher Ind., Inc.). Previous studies on $MgCNi_3$ indicated the need to employ excess magnesium and carbon in the synthesis in order to obtain optimal carbon content[2,16]. The excess Mg is mainly vaporized during the course of the reaction, though MgO is often present in the final product. After thorough mixing, the starting materials were pressed into pellets, wrapped in Zirconium foil, placed on an $Al_2O_3$ boat, and fired in a quartz tube furnace under a 95% Ar / 5% $H_2$ atmosphere. The initial furnace treatment began with a half hour at 600°C, followed by 1 hr at 900°C. After cooling, the samples were reground, pressed into pellets, and placed back in the furnace under identical conditions at 900°C. The latter step was repeated three additional times. Following the final heat treatment, the samples were analyzed with powder X-ray diffraction using $CuK_\alpha$ radiation.

To avoid the problem of neutron absorption of natural abundance B, the samples for neutron diffraction experiments were prepared with 95 % isotopically enriched $^{11}B$. Time-of-flight neutron powder diffraction data were collected on the Special Environment Powder Diffractometer (SEPD)[19] at the Intense Pulsed Neutron Source (IPNS) at Argonne National Laboratory. Diffraction data were collected for samples of nominal composition $MgC_{1.5-x}{}^{11}B_xNi_3$ with x = 0, 0.05, 0.10, 0.15, 0.20, and 0.25 at room temperature. High-resolution backscattering data (2Θ=144.85°, Bank 1) were analyzed using the Rietveld refinement method with the GSAS (EXPGUI) suite[20,21]. Since C and $^{11}B$ neutron scattering cross sections are essentially identical ($0.665 \cdot 10^{-12}$ cm and $0.666 \cdot 10^{-12}$ cm, respectively), only the total occupancy of the (C,B) site in perovskite $Mg(C_{1-x}B_x)Ni_3$ phase was refined. When $MgNi_{2.5}B_2$ was present, its structural parameters were fixed to those obtained in a previous detailed study[22], with phase fraction and cell parameters being the only variables.

**Results and discussion**

A typical Rietveld plot is shown in Fig. 1 for the example of the x=0.15 sample. All the samples of nominal composition $Mg(C_{1.5-x}B_x)Ni_3$ contain MgO and in the samples with x>0.05, the $MgNi_{2.5}B_2$ phase is also observed (Fig. 2). The MgO fraction is around 2% for the whole doping series, and the $MgNi_{2.5}B_2$ fraction grows very slightly with B concentration up to x=0.20, and rapidly increases for x=0.25 from about 1.3% (x=0.2) to 3.5% (x=0.25). At the same concentration, the weight fraction of $MgC_{1-x}{}^{11}B_xNi_3$ drops from 97% (x=0.2) to almost 94% (x=0.25). This indicates that x=0.25 exceeds the boron solubility limit in $MgC_{1-x}{}^{11}B_xNi_3$.

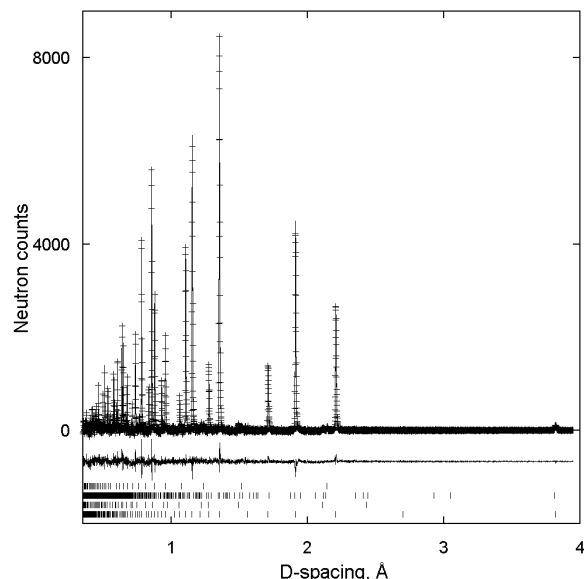

**Fig. 1** Rietveld refinement plot showing the observed (+) and calculated (solid line) diffraction data and their difference for $MgC_{1.5-x}{}^{11}B_xNi_3$, x=0.15 at room temperature. Tick marks, from top to bottom, indicate the intermetallic perovskite phase, MgO, $MgNi_{2.5}B_2$, and vanadium from the sample holder, respectively.

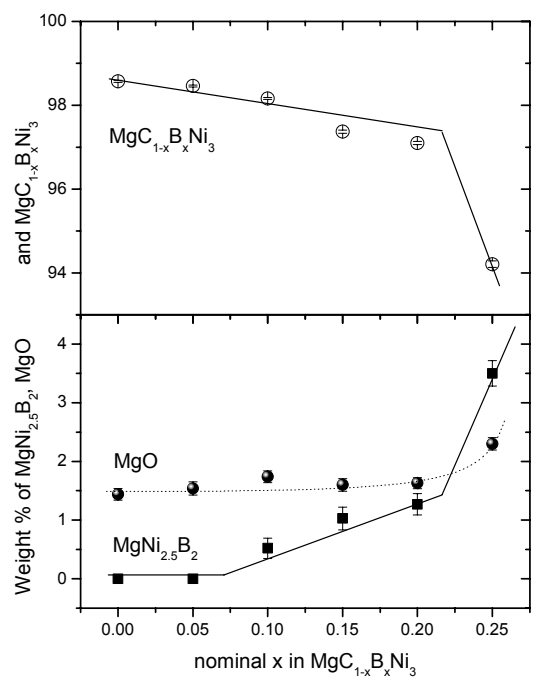

**Fig. 2** The refined weight fractions of the intermetallic perovskite phase, MgO, and $MgNi_{2.5}B_2$ as a function of nominal B content.

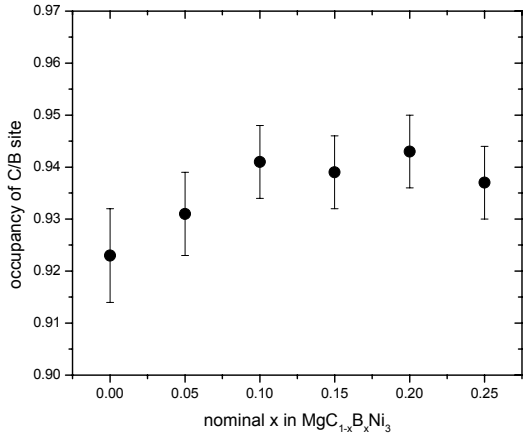

**Fig. 3** Total occupancy of the (C/B) site in $Mg(C_{1-x}B_x)Ni_3$ as a function of nominal boron content.

The C site in this intermetallic perovskite is found to be underoccupied in the whole x range (Fig. 3) in agreement with previous reports[16,23]. To determine the boron doping level, due to the presence of $MgNi_{2.5}B_2$, it was required to show that carbon does not substitute in the boron position in $MgB_2Ni_{2.5}$. Therefore, a series of samples with nominal composition $MgNi_{2.5}B_{2-y}C_y$ (y=0, 0.125, 0.25, 0.375 and 0.50) was prepared. The same synthesis procedure was employed. In Figure 4, the high angle region of x-ray diffraction patterns of all samples is presented. The distinct $\alpha_1$-$\alpha_2$ splitting confirms the high quality of samples. There is no visible shift of the (214), (303), (220) and (206) peaks, indicating that there is no change in the unit cell parameter. Least-squares fits to the 20 strongest X-ray reflections between 20 and 90 degrees 2θ gave the unit cell parameters: a=4.8801(15) Å and 4.8797(11) Å, c=8.788(2) Å and 8.787(2) Å for nominal x=0 and 1.0 respectively.

These are consistent with the reported values a=4.887(2) Å, c=8.789(4) Å (Ref. 22). This difference in cell parameters is less than 1 part in $10^4$, whereas in comparison, the change in cell parameter for 15% B substitution for C in $MgCNi_3$ (see below) is 2 part in $10^3$. Given that carbon and boron have different covalent radii, the lack of change to high precision in the cell parameters in $MgNi_{2.5}B_2$ indicates that there is no solubility of carbon in this phase.

The boron distribution between the intermetallic perovskite phase and $MgNi_{2.5}B_2$ can therefore be specified by Eq. 1:

$MgC_{1-x}B_xNi_3 = MgC_{z-w}B_wNi_3 + (x-w)/2\ MgNi_{2.5}B_2$ (1).

It is known from the refinement that z has about the same value (~0.935) for all samples (Fig. 3), and therefore there is only one unknown, w, in Eq. 1. This value can be found from the experimentally determined weight fractions of the perovskite and $MgNi_{2.5}B_2$ phases.

Figure 5 indicates that the data analysis as described above reveals a linear relationship between the calculated x value and the nominal x in $MgC_{1.5-x}{}^{11}B_xNi_3$ for x<0.25. The last point (x=0.25) does not follow the trend, confirming that x=0.25 exceeds the boron solubility limit in $MgC_{1.5-x}{}^{11}B_xNi_3$. Therefore, further discussion will consider samples only with nominal x < 0.25.

The boron concentration dependence of the lattice constant *a* is shown in Fig.6. For low B contents (x < 0.04) the lattice constant changes slightly as the B doping is increased. For greater x, the lattice expands linearly, from a= 3.81228(2) Å to 3.81966(2) Å for the lowest (x=0.0396) and highest (x=0.155) doping level respectively. This is consistent with the fact that the carbon covalent radius is smaller than that of boron.

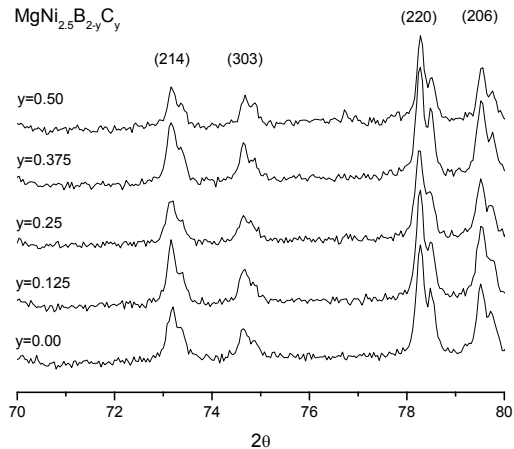

**Fig. 4** Powder x-ray diffraction data (CuK$_\alpha$) for $MgNi_{2.5}B_{2-y}C_y$ for x=0, 0.125, 0.25, 0.375 and 0.5.

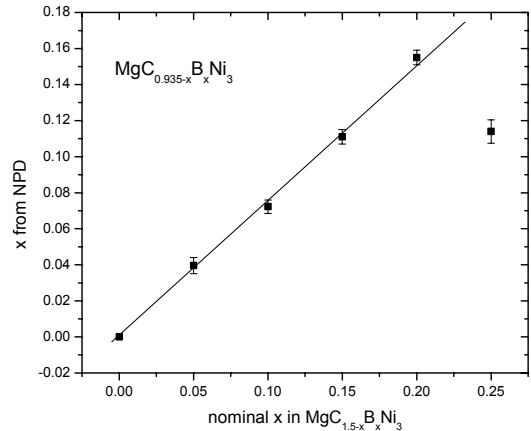

**Fig. 5** Boron content in the intermetallic perovskite phase, determined by neutron diffraction, as a function of nominal boron content.

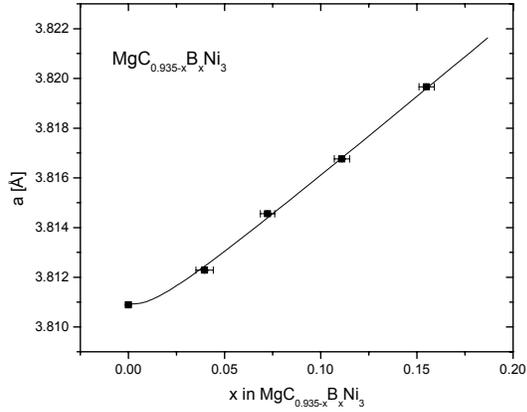

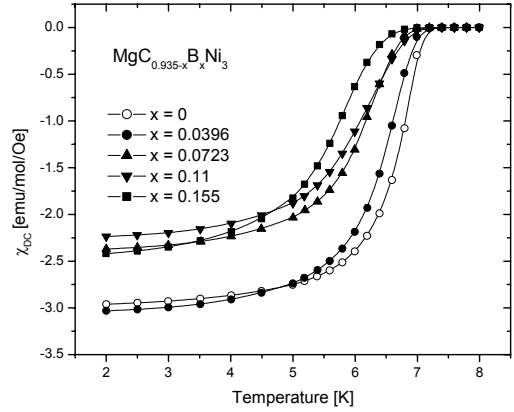

**Fig. 6** The cubic cell parameter as a function of the experimentally determined boron content in the intermetallic perovskite phase. The curve is a guide to the eye.

**Fig. 7** DC magnetization characterization of the superconducting transition for all $MgC_{0.935-x}{}^{11}B_xNi_3$ samples, boron content as determined by neutron diffraction.

Finally, magnetic measurements were made using a commercial SQUID magnetometer (Quantum Design). The superconducting properties were characterized by zero-field cooled DC magnetizations ($H_{DC}$=10 Oe) from 2K to 8K (MPMS – Quantum Design). As seen in Fig.7, in all cases superconductivity remains bulk in character, however smaller diamagnetism is visible for doped samples. The last figure (Fig.8) shows the dependence of the superconducting critical temperature ($T_C$) as a function of boron doping. $T_C$ was determined as the temperature where the extrapolation of the steepest slope of $\chi_{DC}(T)$ in the superconducting state intersects the extrapolation of the normal state to lower temperatures[18]. The x values are those determined in the neutron diffraction experiments. As can be clearly seen in Fig. 8, Tc decreases as boron content decreases: $T_C$ changes from 7.09K to 6.44K going from the undoped sample (x=0) to the highly doped sample (x=0.155).

In the simplest picture, considering calculated DOS, hole doping should increase the DOS at $E_F$ and, as a result, $T_C$ should increase. However, this effect is not observed in the case of partial substitutions on the Ni-site in $MgCNi_{3-x}M_x$ (M=Co, Fe, Ru, Mn)[11-15]. Experiments show that showed that the DOS peak near $E_F$ decreases in cobalt doped and carbon deficient samples[7,10]. More recently, the depression of $T_C$ with increasing carbon deficiency in $MgC_{1-x}Ni_3$ has been proposed to be due to a substantial depression of electron-phonon coupling[17]. Either of these effects could account for the lack of $T_C$ enhancement on hole doping, as observed in these cases and also in the case of $MgC_{1-x}B_xNi_3$.

**Conclusions**

Our neutron powder diffraction studies indicate that the true boron solubility limit in $MgC_{1-x}B_xNi_3$ is 0.16. Magnetic susceptibility measurements show that all samples are superconductors, however superconductivity is suppressed slowly with increasing x. Although boron substitution for carbon is expected to be the least structurally and electronically disruptive of all the chemical substitutions so far successful in $MgCNi_3$, due to the chemical similarity of B and C, this substitution decreases the Tc, suggesting that $MgCNi_3$ has optimal superconducting properties at its intrinsic composition and electron count.

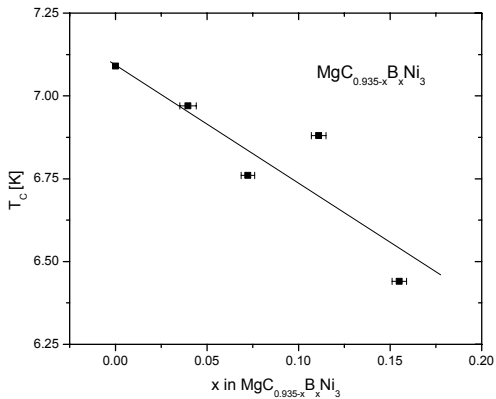

**Fig. 8** Superconducting critical temperature ($T_C$) as a function of x in $MgC_{0.935-x}{}^{11}B_xNi_3$.